\documentclass[10pt,conference,letterpaper]{IEEEtran}
\IEEEoverridecommandlockouts
\usepackage{cite}
\usepackage{amsmath,amssymb,amsfonts}
\usepackage{algorithmic}
\usepackage{graphicx}
\usepackage{textcomp}
\usepackage{xcolor}
\def\BibTeX{{\rm B\kern-.05em{\sc i\kern-.025em b}\kern-.08em
    T\kern-.1667em\lower.7ex\hbox{E}\kern-.125emX}}

\usepackage{soul}

\usepackage{bbm}
\usepackage{graphicx}

\usepackage{amsmath,amssymb,amsfonts}
\usepackage{algorithmic}
\usepackage{graphicx}
\usepackage{textcomp}
\usepackage{xcolor}
\usepackage[ruled,vlined,linesnumbered]{algorithm2e}

\usepackage{mathrsfs}
\usepackage{graphics}
\usepackage{epsfig}
\usepackage{mathptmx}
\usepackage{times}
\usepackage{amssymb}
\usepackage{graphicx}
\usepackage{color}
\usepackage{float}
\usepackage{amsfonts}
\usepackage{amsthm}
\usepackage{indentfirst}
\usepackage{tablefootnote}	
\usepackage{enumerate}
\usepackage{hyperref}
\usepackage{array}
\usepackage{booktabs} 

\usepackage{multirow}
\usepackage{cite}
\usepackage{enumerate}
\usepackage{mathrsfs}
\usepackage{graphicx}
\usepackage{subcaption}

\hypersetup{hidelinks,
 colorlinks=true,
 allcolors=black,
 pdfstartview=Fit,
 urlcolor=magenta,
 breaklinks=true}

\begin{document}

\title{AOC-IDS:  
Autonomous Online Framework with Contrastive Learning for Intrusion Detection

}


\author{
\IEEEauthorblockN{
Xinchen Zhang\IEEEauthorrefmark{2}\IEEEauthorrefmark{3},
Running Zhao\IEEEauthorrefmark{2},
Zhihan Jiang\IEEEauthorrefmark{2},
Zhicong Sun\IEEEauthorrefmark{5},
Yulong Ding\IEEEauthorrefmark{3}, \\
Edith C.H. Ngai\IEEEauthorrefmark{2}\IEEEauthorrefmark{1},
Shuang-Hua Yang\IEEEauthorrefmark{3}\IEEEauthorrefmark{4}\IEEEauthorrefmark{1}
}

\IEEEauthorblockA{
\IEEEauthorrefmark{2}The University of Hong Kong
\IEEEauthorrefmark{3}Shenzhen Key Laboratory of Safety and Security for Next \\ Generation of Industrial Internet, Southern University of Science and Technology}
\IEEEauthorblockA{
\IEEEauthorrefmark{4}University of Reading
\IEEEauthorrefmark{5}The Hong Kong Polytechnic University
}

\thanks{\IEEEauthorrefmark{1}Co-corresponding authors: Edith C.H. Ngai (Email: chngai@eee.hku.hk), Shuang-Hua Yang (Email: shuang-hua.yang@reading.ac.uk)}
}

\maketitle
\begin{abstract}
The rapid expansion of the Internet of Things (IoT) has raised increasing concern about targeted cyber attacks. Previous research primarily focused on static Intrusion Detection Systems (IDSs), which employ offline training to safeguard IoT systems. However, such static IDSs struggle with real-world scenarios where IoT system behaviors and attack strategies can undergo rapid evolution, necessitating dynamic and adaptable IDSs. In response to this challenge, we propose AOC-IDS, a novel online IDS that features an autonomous anomaly detection module (ADM) and a labor-free online framework for continual adaptation. In order to enhance data comprehension, the ADM employs an Autoencoder (AE) with a tailored Cluster Repelling Contrastive (CRC) loss function to generate distinctive representation from limited or incrementally incoming data in the online setting. Moreover, to reduce the burden of manual labeling, our online framework leverages pseudo-labels automatically generated from the decision-making process in the ADM to facilitate periodic updates of the ADM. The elimination of human intervention for labeling and decision-making boosts the system's compatibility and adaptability in the online setting to remain synchronized with dynamic environments. Experimental validation using the NSL-KDD and UNSW-NB15 datasets demonstrates the superior performance and adaptability of AOC-IDS, surpassing the state-of-the-art solutions. 
The code is released at \href{https://github.com/xinchen930/AOC-IDS}{https://github.com/xinchen930/AOC-IDS}.
\end{abstract}



\section{Introduction}
The exponential growth of the Internet of Things (IoT) has revolutionized multiple sectors, including transportation, healthcare, smart cities, agriculture, and more \cite{POURRAHMANI2022100579,8767247,8124196,6740844,9374808
}. However, it also escalates cybersecurity threats. For example, the malware attacks targeted at IoT systems aggravate every year\footnote{https://www.statista.com/statistics/1322216/worldwide-internet-of-things-attacks/}. 
Given this, developing Intrusion Detection Systems (IDS) that can adapt to evolving or new attacks and efficiently learn from enormous data is paramount \cite{GARCIATEODORO200918,FERRAG2020102419}. The advances in machine learning, particularly deep learning, have boosted this development \cite{5687239,8369054,8126009,8066291
}.

Basically, IDS falls into two categories: signature-based detection and anomaly-based detection \cite{GARCIATEODORO200918
}. While signature-based detection efficiently identifies known threats by comparing data against a library of threat signatures, it struggles with unknown threats \cite{GARCIATEODORO200918
}. Conversely, anomaly-based detection systems depend on their comprehension of regular behavioral patterns, termed as `representations' or `features'. They trigger alerts upon identifying deviations from these established patterns, thereby facilitating the detection of novel or `zero-day' attacks \cite{al2020anomaly}. This approach provides a significant advantage over signature-based detection, particularly within an ever-evolving threat landscape. Therefore, we concentrate on anomaly-based systems to handle new and changing threats.


In dynamic open-world scenarios, not only new malicious system patterns like zero-day attacks can emerge, but also benign behavior of the system can evolve over time. This evolution could be attributed to environmental changes or shifts in user preferences, both capable of altering normal patterns. Given the reliance of anomaly-based intrusion detection on accurate modeling of normal system patterns, any shift in the normal-abnormal behavior boundary could impair detection performance.
Therefore, anomaly-based intrusion systems necessitate continuous adaptation to the evolving normal behavior of the system over time. 
Online learning offers an effective solution to enhance the system's adaptability through ongoing monitoring and updating system behavior profiles. However, there are still significant research gaps in the application of online learning to anomaly-based IDS, especially for identifying anomalies from streaming data in IoT systems. Most existing learning-based anomaly detection systems train a static model offline, lacking dynamism in an ever-changing environment. It is important to note that many existing works related to `online IDS' often refer to IDS that detects intrusions by performing real-time inferences, in which its training process remains the same as that of offline IDS with no further model updates \cite{9837465,BALDINI2022108923,https://doi.org/10.1049/cps2.12016
}.


Establishing an effective online learning framework for intrusion detection necessitates prompt adaptation of the IDS to a continuous influx of unlabeled data. This situation gives rise to a number of significant challenges.

Primarily, online learning requires the model to establish a normal profile of system patterns from limited or incrementally incoming data and be able to adapt to changing data distributions without extensive retraining. To increase the representation learning capability of the model, we employ contrastive learning \cite{InfoNCE
}, a method renowned for its ability to learn robust and informative representations by attracting positive (similar) pairs and repelling negative (dissimilar) pairs, enabling the anomaly-based IDS to capture common characteristics of benign behaviors effectively and distinguish these benign representations from malicious ones.
Specifically, we propose a novel supervised contrastive loss, named Cluster Repelling Contrastive (CRC) loss. The CRC loss, coupled with the simultaneous use of both encoder and decoder outputs from an Autoencoder (AE), significantly enhances the representation capability of the detection model with limited available information. 
As for the post-processing of data representations, we implement an \emph{autonomous} statistical decision-making process based on Gaussian distributions for intrusion detection. Compared to traditional methods, which usually require manual threshold selection, the proposed decision-making method is more compatible with the online setting. 

Moreover, online learning also confronts the inherent challenge of effectively utilizing substantial volumes of unlabeled data. 
Many existing approaches \cite{pendlebury2019tesseract,jan2020throwing} employ a straightforward method of manually labeling new data, which is a labor-intensive process that compromises the practicality of the system.
To address this issue, we propose the use of pseudo-labels generated by the IDS itself based on current knowledge to complete the intrusion detection process and the dataset expansion \emph{autonomously}. Leveraging these pseudo-labels facilitates the use of supervised contrastive loss, which provides clearer guidance for the learning algorithm than unsupervised ones. 
This strategy significantly reduces the labor for labeling a high volume of unlabeled data and ensures model performance through a supervised contrastive loss.

In this work, our contributions are summarized as follows:

\begin{itemize}
    \item We present AOC-IDS, an autonomous online IDS, featuring a distinct anomaly detection module (ADM) and an innovative online framework enabling continual adaptation.
    Our system design delivers 
    a robust solution for intrusion detection in dynamic environments where the system behaviors and attack strategies evolve over time.

    \item We employ an AE with a custom-designed CRC loss function and leverage both encoder and decoder outputs for superior data comprehension and enhanced representation capability within the ADM. The utilization of contrastive learning benefits the discrimination of representations for differentiating malicious behavior based on normal system patterns.
    
    \item In the online learning framework, we utilize an \emph{autonomous} decision-making process without human intervention to generate pseudo-labels. These pseudo-labels alleviate the demand for manual labeling and facilitate \emph{autonomous} updates, 
    making our ADM highly compatible with the online framework.
    
    
    \item We conduct extensive experiments on network traffic datasets, NSL-KDD and UNSW-NB15 datasets, to demonstrate the optimal performance and adaptability of AOC-IDS. The ablation study reveals the contributions of individual components.


    
    

\end{itemize}

\section{System Overview}
Developing an online learning-based IDS that can effectively adapt to dynamic environments involves two key components: 
(\romannumeral 1) Anomaly Detection Module: This necessitates the establishment of an ADM that is proficient in detecting intrusions, especially for the online setting with limited or incrementally incoming data.
(\romannumeral 2) Online Framework: This calls for the effective utilization of unlabeled new data to perpetually update the ADM.

\begin{figure}[!t]
    \centering
    \includegraphics[width=0.5\textwidth]{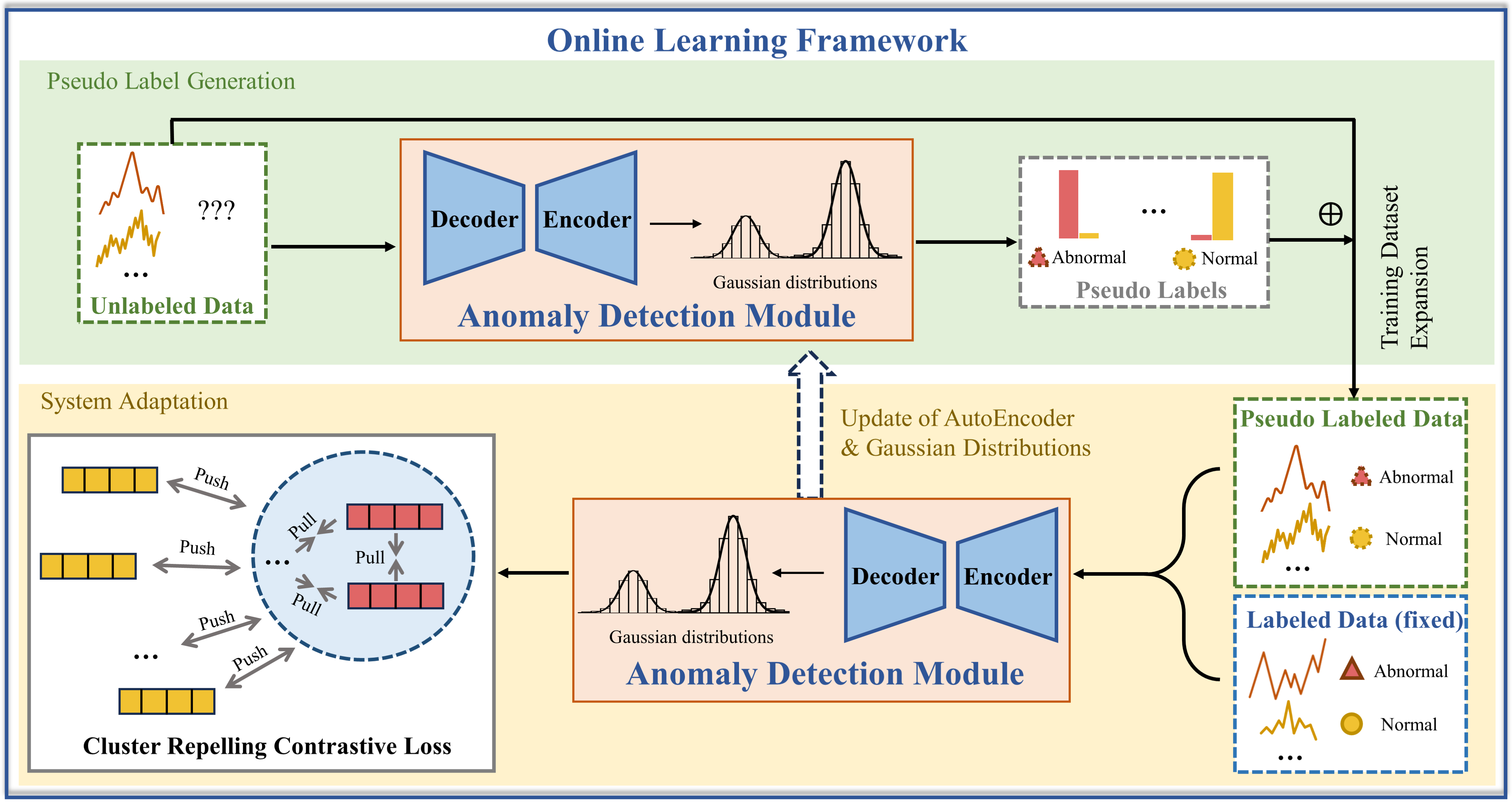}
    \caption{System overview of AOC-IDS. In the proposed system, the ADM undergoes adaptation within the online framework. The online framework consists of two steps: pseudo-label generation and system adaptation. The ADM extracts the feature of an input using an AE and the extracted feature is labeled according to the Gaussian fit result.}
    \label{fig:training}
    \vspace{-0.5cm}
\end{figure}

\figurename{~\ref{fig:training}} gives an overview of AOC-IDS. The ADM extracts the feature of an input using an AE and the extracted feature is labeled according to the Gaussian fit result. The ADM also undergoes adaptation within the online framework. The online framework consists of two steps, pseudo-label generation and system adaptation, to continually update the ADM.




\subsection{Anomaly Detection Module}
The ADM employs a two-step approach, including representation learning and decision-making, to establish a robust, fully automated intrusion detection mechanism.

\textbf{Representation Learning Process.} The first process in the ADM focuses on learning distinctive and informative representations from the input data. An AE with a custom-designed contrastive loss function is proposed to utilize the available attack information to learn more discriminative data representations. The AE utilizes the outputs of both the encoder and the decoder to incorporate all available information. The tailored contrastive loss groups similar
samples together and distances dissimilar ones. 
The combination of them enhances the model's learning capability and ensures a more comprehensive understanding of the input data.

\textbf{Decision-Making Process.} The second process in the ADM identifies potential intrusions based on the learned data representations. Unlike conventional methods that rely on manually predefined thresholds, our system uses a statistical approach to detect potential threats autonomously. Specifically, we fit the similarity distribution of the learned representations into two Gaussian distributions, one for normal patterns and the other for abnormal ones. For an input waiting to be labeled, we employ the statistical confidence of its association with these two distributions to determine its integrity, thereby facilitating autonomous intrusion detection.

\subsection{Online Framework}

The online framework updates the ADM and dataset continually, preserving the system's adaptability to evolving threats. It has two phases: \textbf{(\romannumeral 1) pseudo-label generation} based on inference from the ADM, eliminating manual labeling for autonomous updates, and \textbf{(\romannumeral 2) system adaptation} for model fine-tuning and the regeneration of Gaussian distribution parameters aligned with the continuously updated training dataset.

Initially, a detection model is trained using a small labeled dataset. Once a certain amount of new inputs has been pseudo-labeled, the model is updated using the expanded dataset, including both true (initial) labels and pseudo-labels.

Our framework incorporates specific considerations, such as the introduction of random noise and constraints on the acquisition of Gaussian distributions. These factors alleviate potential overfitting due to incorrect judgments, thereby enhancing the overall robustness of the online learning framework.




\begin{figure}[!t]
    \centering
    \includegraphics[width=0.5\textwidth]{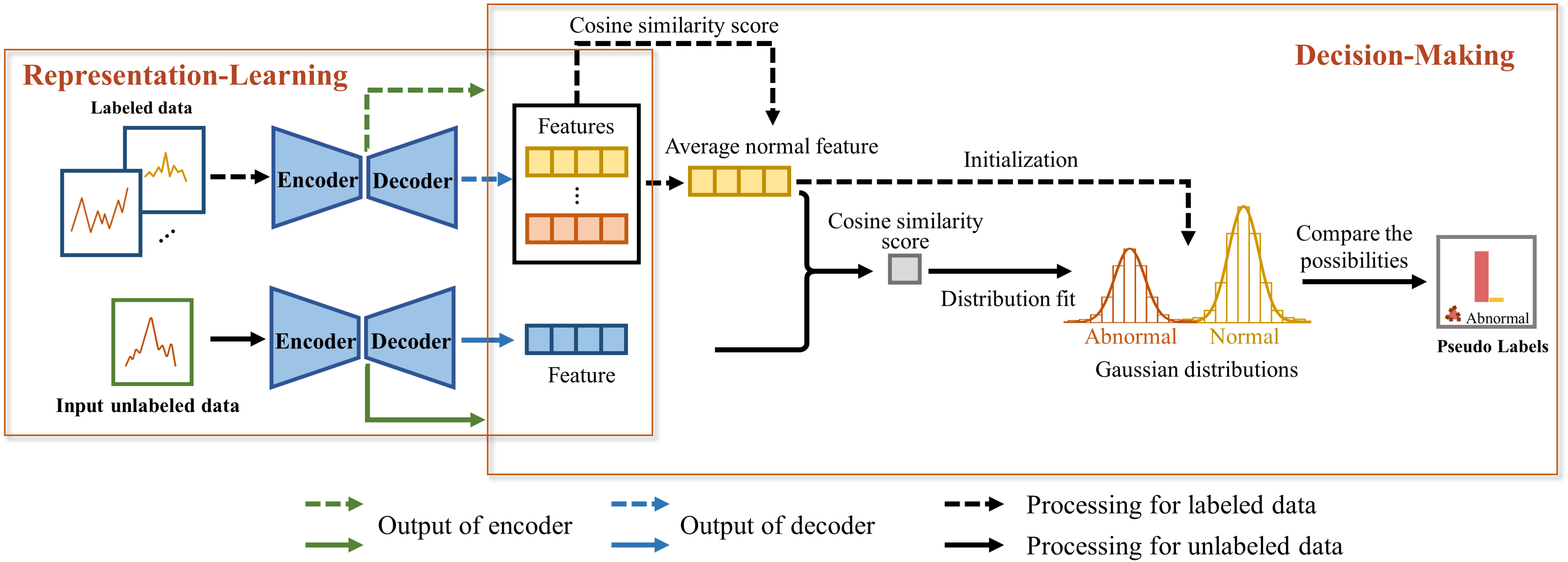}
    \caption{Inference process of the proposed ADM to label an input.}
    \label{fig:workflow}
    \vspace{-0.5cm}
\end{figure}

\section{Our Intrusion Detection System}
\label{sec:ids}

This section details the design and operation of our proposed IDS, comprising the ADM and the online framework. The ADM's role is to detect intrusions and generate pseudo labels. Meanwhile, the online framework continuously updates the ADM with incoming unlabeled data, autonomously. This combination enables proficient and adaptive intrusion detection.



\subsection{Anomaly Detection Module}

    In cybersecurity, the efficacy of an anomaly-based detection system depends on its understanding of normal behavioral patterns. It is pivotal for the system to generate discriminative representations and discern benign representations from potential threats for accurate intrusion detection.

    As depicted in \figurename{~\ref{fig:workflow}}, the ADM thus comprises a representation-learning process, responsible for the extraction of salient features from the input and training data, and a decision-making process, which established the normal behavioral patterns and separates potential malicious features from normal behavioral patterns autonomously.  
    
    \textbf{Representation Learning with a Novel Contrastive Loss.} 
    We propose a novel AE incorporating supervised contrastive learning to generate more discriminative representations for intrusion detection from both the encoder and decoder, enhancing the differentiation between normal and abnormal patterns.

    Our AE, while maintaining a similar structure to traditional AEs, features a modified CRC loss and focuses on extracting salient intrusion detection features from input vectors via both the encoder and decoder. The AE is trained to generate similar representations for all normal instances, while distancing the intrusion representations. This method is more straightforward than reconstructing normal instances for intrusion detection. Consequently, the proposed approach leads to a more efficient and accurate intrusion detection model.

    In contrastive learning, an anchor sample is a reference data point used for comparison. During training, similar examples (positive samples) are encouraged to have high similarity, and dissimilar examples (negative samples) are encouraged to have low similarity with the anchor. This process aids the model in learning discriminative representations for grouping similar samples together and distancing dissimilar ones.


    In intrusion detection, the variability of zero-day attacks necessitates a training focus on normal behavior, which is consistent, unlike the divergent patterns of attacks. Hence, our contrastive learning uses only normal samples as anchors to construct positive pairs, treating all abnormalities as negatives.
    

    
    \noindent \emph{Notations.} In the subsequent analysis, the subscripts `$n$' and `$a$' represent `normal' and `abnormal', while the superscripts `$en$' and `$de$' represent the encoder and decoder. For a fixed training dataset $D$ inclusive of both normal and abnormal data:
    \begin{align*}
    D &= \{D_n, D_a\} = \{X_n, Y_n, X_a, Y_a\}\\
    &= \{\{x_1, ..., x_{l_n}\}, \{y_1, ..., y_{l_n}\}, \{x_1, ..., x_{l_a}\}, \{y_1, ..., y_{l_a}\}\},
    \end{align*}
    where $X$ and $Y$ stand for the set of input vectors and their corresponding labels. The dataset comprises $l_n$ ($\in \mathbb{N}$) normal inputs and $l_a$ ($\in \mathbb{N}$) abnormal inputs. Each input vector $x$ has a dimension $d$ ($\in \mathbb{N}$), while each label $y \in \{0, 1\}$, where 0 signifies normal, and 1 signifies abnormal. We denote each learned representation of the input in the training dataset as $v \in V$. The detection model in our work is denoted as $AE_{\theta}$, where $\theta$ is the model weight. 

    In this work, we propose the CRC loss function extended from the standard InfoNCE \cite{InfoNCE}, one of the most popular losses in contrastive learning. 
    For an anchor sample representation $v_{n,i}$ from the normal class, the goal is to maximize the similarity between the positive pair \{$v_{n,i}, v_{n,j}$\}, where $i,j \in \{1,2,...,l_n\}$, while minimizing the similarity between the anchor sample $v_{n,i}$ and negative example set $\{v_{a,k} | k \in \{1,2,...,l_a\}\}$. 
    The CRC loss denoted by $\mathcal{L}$, can then be formulated as:
    \begin{equation}
        \label{eq:Lij_info}
        \mathcal{L}_{ij} = -log \frac{exp(h(i,j)/\tau)}{exp(h(i,j)/\tau) + \sum_{i=1}^{l_n} \sum_{k=1}^{l_a} exp(h(i,k)/\tau)},
    \end{equation}
    \begin{equation}
        \label{eq:L}
        \mathcal{L} = \frac{1}{l_n (l_n - 1)} \sum_{i=1}^{l_n} \sum_{j=1,j \neq i}^{l_n} \mathcal{L}_{ij},
    \end{equation}
    where $\tau \in (0,1]$ is the temperature parameter and the function $h(\cdot)$ is a function that measures the similarity between two vectors, detailed in (\ref{eq:cos}).

\begin{figure}[!t]
    \centering    \includegraphics[width=.5\textwidth]{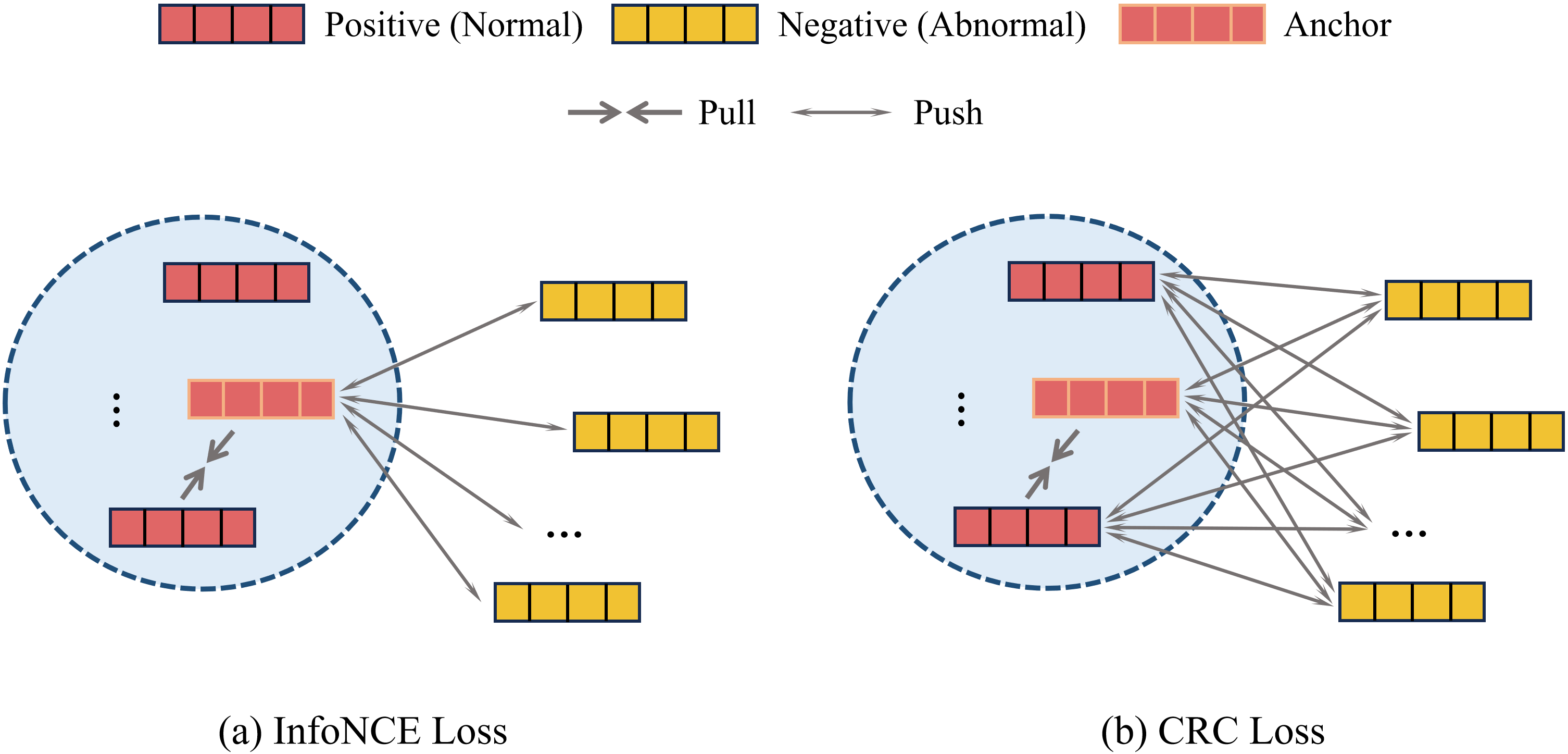}
    \caption{Difference between InfoNCE loss and CRC loss in intrusion detection.}
    \label{fig:infonce_and_CRC}
    \vspace{-0.5cm}
\end{figure}
    
    It is worth noting that when selecting the negative sample pairs, we traverse not only all the negative samples corresponding to each fixed anchor sample but also all the available anchors in the normal class. This adjustment enables a dual-category repulsion effect at each stage, unlike the traditional method where only a single anchor repels negative samples and does not pay attention to the relationship between negative examples and other positive samples, demonstrated in \figurename{~\ref{fig:infonce_and_CRC}}.

    In this work, we use the cosine similarity score to evaluate the similarity between two vectors, defined as follows: 
    \begin{equation}
        \label{eq:cos}
        h(i,j) = CosSim(v_i, v_j) = \frac{v_i^\top v_j}{||v_i|| ||v_j||}.
    \end{equation}
    
    Notably, the CRC loss is applied to both the encoder and decoder to learn representations.
    The final loss function, $\mathcal{L}^{final}$, is the sum of the losses from the encoder ($\mathcal{L}^{en}$) and the decoder ($\mathcal{L}^{de}$):
    \begin{equation}
        \label{eq:L_all}
        \mathcal{L}^{final} = \mathcal{L}^{en} + \mathcal{L}^{de}.
    \end{equation}

    By minimizing the loss function, we can achieve a dual objective: to simultaneously maximize the similarity among normal data representations and minimize the similarity between normal representations and those of intrusions.

    \textbf{Autonomous Decision-Making Process.} Our ADM incorporates an autonomous decision-making process, as detailed in \figurename{~\ref{fig:workflow}}. It starts with the learned representations of all training data $D$ and the average (mean) representation of normal training data $D_n$. 
    
    Subsequently, we calculate cosine similarity scores between the average normal representation and all training dataset representations. Intuitively, the distribution of these scores should follow the combined form of two Gaussian distributions. The Gaussian distribution with a higher mean parameter should represent the `normal' distribution because normal representations should have greater similarity to the average normal representations by nature. In contrast, the Gaussian distribution with a lower mean represents the `abnormal' distribution.

    
    To formalize this intuition, we use Maximum Likelihood Estimation (MLE) to fit cosine similarity scores into two Gaussian distributions, denoted as `normal' and `abnormal'. This label-free optimization reduces overfitting risks on challenging samples.
    
    


    During inference, after obtaining the representation of each test input from AE, we calculate the cosine similarity score between the test representation and the average normal representation. This score is then applied to both normal and abnormal Gaussian distributions to derive the associated probabilities. The class of the test data corresponds to the distribution type with the higher probability.


    Our detection model independently processes input vectors via the encoder and decoder to generate representations. 
    As a result, the procedure mentioned above applies to both the encoder and decoder. The final decision stems from a voting mechanism that considers the outputs from both of them. The probability of the input correlating with each distribution serves as a confidence measure of the results, with the outcome exhibiting higher confidence elected as the final decision.
    
    

    \begin{figure}[!t]
        \centering    
        \includegraphics[width=.5\textwidth]{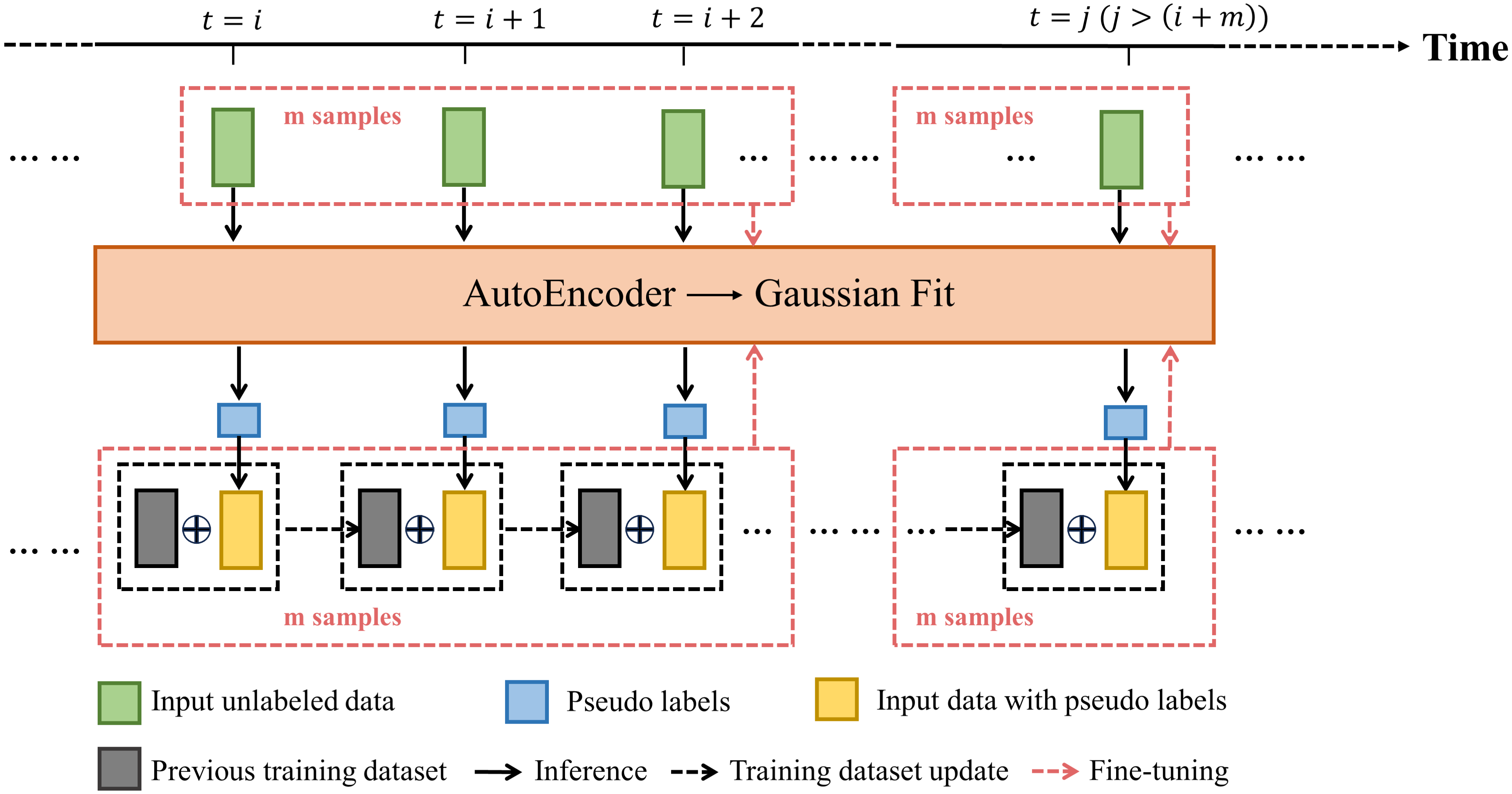}
        \caption{Timeline of the proposed online framework.}
        \label{fig:timeline}
        \vspace{-0.5cm}
    \end{figure}

\subsection{Incorporating Online Learning}
    
    In the traditional closed-world scenario, one assumes that training data capture every possible instance or situation the model may encounter post-deployment, which an offline-trained static IDS handles successfully. However, this assumption often falls short in real-world applications where environmental conditions and system patterns are dynamic, necessitating continuous adaptations of the detection system.
        
    
    We propose an online learning framework that enhances our system's adaptability to continuously changing environments. 
    As shown in \figurename{~\ref{fig:timeline}}, our online framework follows a sequence that includes expanding the training dataset with pseudo-labels and periodically fine-tuning the ADM based on this expanded dataset. 
    A salient feature of our approach is its independence from manual labeling during the online training process, thereby bolstering the system's practicality and efficiency. 
    
    
    
\begin{algorithm}[t]
    \caption{Framework of online training.}
    \label{alg:online}
    \SetAlgoLined
    \SetKwFunction{Train}{Train}
    \SetKwFunction{ID}{IntrusionDetection}
    \SetKwFunction{Fit}{FitGaussian}
    \SetKwFunction{DC}{DecisionMaking}
    \SetKwFunction{Vote}{Vote}
    \SetKwFunction{flip}{RandomFlip}
    
    \KwData{Training dataset for the first round of training: $D_0 = \{X_0, Y_0\}$\; 
        \setlength{\parindent}{2.7em} Streaming new input vectors: $\{x_{l_0+1},..., x_i, ...\}$, $i > l_0, i \in \mathbb{N}$.
    }
    \BlankLine
 
    Initialize the model weight $\theta$ randomly\;
        \Train{$X_0, Y_0, AE_{\theta}, epoch_0$}; \tcp*[f]{\textnormal{train the model $AE_{\theta}$ for $epoch_0$ rounds on initial dataset $D_0$}}\;
        $X, Y = X_0, Y_0$\;
        
    \While{\textnormal{inputting} $x_i$}{
            $G^{en}_n, G^{en}_a, G^{de}_n, G^{de}_a$ = \Fit{$X_0, X, AE_{\theta}$};
            \tcp*[f]{\textcolor{gray}{\textnormal{obtain Gaussian distributions of encoder and decoder's based on initial dataset;}}}\
            
        \For{$j = 0$ \textnormal{to} $m-1$}{
            $u^{en}_{i+j}, u^{de}_{i+j} = $ $AE_{\theta}(x_{i+j})$; \tcp*[f]{\textcolor{gray}{\textnormal{obtain both encoder and decoder's outputs of $x$;}}}\
                
            $\hat{y}_{i+j}^{en}, \hat{y}_{i+j}^{de} = $ \DC{$u^{en}_{i+j}, G^{en}_n, G^{en}_a$}, \DC{$u^{de}_{i+j}, G^{de}_n, G^{de}_a$} ; \tcp*[f]{\textcolor{gray}{\textnormal{obtain pseudo-labels given by encoder and decoder respectively;}}}\

            $\hat{y}_{i+j} = $ \Vote{$\hat{y}_{i+j}^{en}, \hat{y}_{i+j}^{de}$}\; \tcp*[f]{\textcolor{gray}{\textnormal{select the result with higher confidence; }}}\
                
            $\{\hat{y}_{i+j}\} = $ \flip{$\{\hat{y}_{i+j}\}, \lambda$}; \tcp*[f]{\textcolor{gray}{\textnormal{randomly flip $\lambda$ pseudo-labels;}}}\ 
            
            Update $X, Y = X \cup \{x_{i+j}\}, Y \cup \{\hat{y}_{i+j}\}$\;
        }

            \Train{$X, Y, AE_{\theta}, epoch_1$}; \tcp*[f]{\textcolor{gray}{\textnormal{train the model $AE_{\theta}$ for $epoch_1$ rounds on updated dataset $\{X,Y\}$;}}}\
    }
\end{algorithm}

    The online learning framework is presented in Algorithm \ref{alg:online}. 
    Initially, the detection model $AE_{\theta}$ with random initialized weight $\theta$ is trained on a small initial labeled training dataset $D_0 = \{X_0, Y_0\}$ for $epoch_0$ rounds (Algorithm \ref{alg:online} line 2).
    Subsequently, the online training process involves two stages: the pseudo-label generation phase (Algorithm \ref{alg:online} lines 5-9), and the system adaptation phase (Algorithm \ref{alg:online} lines 10-13).
    
    \textbf{Pseudo-label Generation.} In the pseudo-label generation phase, we assign a `pseudo-label' to each new input vector $x_j$ passing through the ADM by the current model $AE_{\theta}$. If the input is identified as anomalous (i.e., the pseudo-label $\hat{y}_j$ is 1), the system promptly triggers an alarm. During the pseudo-label generation phase, the ADM performs inference to generate predictions, thereby fulfilling the intrusion detection task.
    
    
    \textbf{System Adaptation.} In the system adaptation phase, each new input $x_j$ and its corresponding pseudo-label $\hat{y}_j$ are incorporated to expand the training dataset $D = \{X, Y\}$. After detecting $m$ new data entries, we fine-tune the current detection model $AE_{\theta}$ over $epoch_1$ iterations on the updated training set $D$. It is worth noting that the updated label set $Y$ includes both true labels and pseudo-labels.
    
    Nonetheless, the expanded training dataset may include mislabeled data, potentially causing the model to deviate from the truth and leading to inferior performance compared to a scenario where no updates were made.
    
    Therefore, to resist the adverse effect of false pseudo-labels during system updates, our framework incorporates specific strategies, including constraints on the acquisition of Gaussian distributions and the introduction of random noise 
    to enhance the robustness of the online learning framework.  
    
    More specifically, we recognize the crucial role of the average normal representation in generating distributions. Thus, we restrict the averaging of normal representations to the initial training dataset $X_0$, which consists solely of true labels. This restriction (Algorithm \ref{alg:online} line 5) ensures the reliability of the acquired Gaussian distributions. 
    The random noise strategy (Algorithm \ref{alg:online} line 10) effectively mitigates the issue of overfitting one's own wrong judgments in a self-learning context. It accomplishes this by diversifying the wrong judgments from a singular `direction' to various divergent `directions', thereby preventing the model from continuously reinforcing its own mistaken interpretations.

    Furthermore, we designed the ADM with the compatibility of the online framework in mind. The autonomous decision-making process, which requires minimal label information, endows the system with a high degree of resistance to false label information. This is due to the Gaussian fit process merely focusing on the shape of the distribution, rather than the class affiliation of the points within it.

    Through the implementation of these strategies, we can create a robust system that consistently delivers high performance, demonstrating resistance to inaccuracies from pseudo-labels.

\section{Experiments}
\label{sec:experiments}

This section presents the experimental setup and results. We detail the datasets, experiment settings, and baseline methods in \subsectionautorefname{~\ref{sec:dataset}}. The superior performance of AOC-IDS over the baseline methods in the online setting is demonstrated in \subsectionautorefname{~\ref{sec:comparative experiments}}. Additionally, ablation experiments are conducted to illustrate the contribution of sub-components in AOC-IDS in \subsectionautorefname{~\ref{sec:ablation experiments}}. 
The experimental results reveal that our method outperforms the state-of-the-art (SOTA) techniques and underscores the significant contributions of each component in our system to its capacity and adaptability.

\subsection{Dataset Preparation and Experiment Settings}
\label{sec:dataset}

\subsubsection{Datasets}

    To verify the effectiveness of the proposed method in intrusion detection, we conduct experiments on two datasets, NSL-KDD and UNSW-NB15 \cite{FeCo,conIDS,pajouh2017two
    }, which are widely used in the field of IoT system intrusion detection.
    
    \textbf{NSL-KDD Dataset}. A standard benchmark in intrusion detection, NSL-KDD comprises 125,973 training and 22,544 test network traffic samples, with 24 attack types in training and 38 in testing. These attacks span four categories: DoS (Denial of Service Attack), U2R (User to Root Attack), R2L (Remote to Local Attack), and Probing Attack. It is notable that some attack types only exist in the test set rather than the training set, which makes the tasks closer to real-world scenarios, i.e., zero-day attack detection. 
    
    
    \textbf{UNSW-NB15 Dataset.} UNSW-NB15 contains a hybrid of the real modern norm and the contemporary synthesized attack activities of the network traffic. To comprehensively represent the diversity of attacks, nine attack types are considered in the dataset, including Fuzzers, Analysis, Backdoors, DoS, Exploits, Generic, Reconnaissance, Shellcode, and Worms. This dataset is divided into the training set with 175,341 samples and the test set with 82,332 samples. 

    \subsubsection{Baselines}
\label{sec:baselines}

    We compare the proposed method with the following SOTA baseline IDSs \cite{FeCo,conIDS} and some other widely-used machine learning baselines, including decision tree classifier (DTC), random forest (RF), support vector machine (SVM), and XGBoost, in the same online setting.
    
    \begin{itemize}
        \item \textbf{FeCo} \cite{FeCo} defines a contrastive loss, which is in the form of InfoNCE, for maximizing the distance between benign and malicious representations while minimizing the distance among benign representations. The learned representations are fed into a decision-making process with a fixed threshold. 
        \item \textbf{CIDS} \cite{conIDS} is a contrastive learning-based IDS with a contrastive cross-entropy loss which is a combination of contrastive loss and classification loss. The labels can be directly generated by the detection model. 
    \end{itemize}


    \subsubsection{Experiment Settings} 
    Our pre-processing approach is straightforward and avoids complex feature engineering. It includes two steps. Firstly, we remove any attribute with the same value across the entire dataset as it does not contribute to the task. In the NSL-KDD dataset, we removed one attribute with a constant value, reducing the dimensionality of training attributes from 41 to 40. The UNSW-NB15 dataset had no such attributes. Therefore, all 42 attributes were retained. Next, we normalize the continuous attributes and apply one-hot encoding to discrete ones. Following these steps, the NSL-KDD dataset ends up with 121 dimensions, while the UNSW-NB15 dataset with 196 dimensions. 
    The data pre-processing procedures are applied consistently across all comparative methods to facilitate an equitable performance comparison.

    The experiments are, by default, conducted in an online setting, a scenario in which the IDS initially has access only to a limited number of labeled datasets. The remaining data is gradually fed into the system in a streamlined manner.

    For both datasets, the ADM initially trains on 20\% of the original training datasets and undergoes an update after every incremental addition of 1.6\% of the original training dataset (2000 samples for NSL-KDD, and 2784 samples for the UNSW-NB15 dataset). 
    For the NSL-KDD dataset, \(epoch_0 = 4\), \(epoch_1 = 1\), and \(\lambda = 20\%\). 
    For the UNSW-NB15 dataset, \(epoch_0 = 300\), \(epoch_1 = 3\), and \(\lambda = 5\%\). 


    The architecture of our AE comprises layer sizes [121, 64, 32, 64, 121] for NSL-KDD and [196, 128, 64, 128, 196] for UNSW-NB15. The training process employs a Stochastic Gradient Descent (SGD) optimizer with a learning rate of 0.001 and a batch size of 128. The temperature \(\tau\) in the contrastive loss function is set to 0.02.

    The reported results are the average of five consecutive rounds of experimentation, guaranteeing their reliability.



\begin{table}[!t]
\caption{Performance (\%) comparison between AOC-IDS and baselines. The highest metric performance is bolded.}
\resizebox{\columnwidth}{!}{%
\begin{tabular}{ccccccccc}
\toprule
\multirow{2}{*}{Method} & \multicolumn{4}{c}{NSL-KDD} & \multicolumn{4}{c}{UNSW-NB15} \\ \cmidrule(lr){2-5} \cmidrule(lr){6-9}
                        & Acc.   & Pre.  & Rec.  & F1  & Acc.    & Pre.   & Rec.   & F1   \\ \midrule
\textbf{AOC-IDS}        & \textbf{88.90}  &85.99  &\textbf{96.21} &   \textbf{90.81} &\textbf{89.19}  & 90.65   & 89.70    &\textbf{90.14}      \\
FeCo       &   81.10     &  90.72     &  74.44     & 81.72 & 72.50   &   \textbf{91.18}     &  55.41     &  68.93     \\
CIDS      &   82.29     &  93.54     &   77.18    &  80.65   &   82.61   &   78.91    & \textbf{96.28}      &   86.03      \\
DTC                      &  78.98 &  92.27  &  69.28  &  78.92 &  85.95 &  82.32  &  94.86  &  88.15    \\
RF                      &  76.74  &  \textbf{96.71}  &  61.23  &  74.99 &  85.93 &  80.25  &   98.75  &  88.55  \\
SVM                     &  75.60  &  91.87  &  62.69  &  74.52 &  56.44  & 56.00  & 97.98  &  71.24    \\
XGBoost                 &   77.95 &  96.46  &  63.60  &  76.62 &    86.97    &   82
85&   98.29    &  89.26    \\ \bottomrule
\end{tabular}}
\label{tab:compare}
\vspace{-0.5cm}
\end{table}

\subsection{Performance of AOC-IDS}
\label{sec:comparative experiments}
    We undertake a comparative evaluation of AOC-IDS against both SOTA deep learning methodologies and well-established machine learning methods, as detailed in \subsubsectionautorefname{~\ref{sec:baselines}}. This comparison is conducted within the same online setting to highlight the efficacy and adaptability of our system with limited labeled data. Moreover, we underscore the exceptional capacity of AOC-IDS for detecting zero-day attacks in dynamic environments through an analysis of the detection rates of zero-day attacks on the NSL-KDD test dataset.
    
    \subsubsection{Overall Performance}
        We present the comparative evaluation result based on the mean values of accuracy, precision, recall, and F1 scores over five consecutive rounds. While precision and recall exclusively focus on the performance of the positive class and therefore cannot fully represent the performance of the system, accuracy and the F1 score typically offer a more comprehensive performance evaluation.
        Consequently, we rely primarily on accuracy and the F1 score to assess system performance in subsequent discussions.
        
        As shown in \tablename{~\ref{tab:compare}}, AOC-IDS outperforms other methods in terms of accuracy and F1 score on both NSL-KDD and UNSW-NB15 datasets. AOC-IDS achieves detection accuracy as high as 88.90\% and 89.19\% on NSL-KDD and UNSW-NB15 datasets with about 7\% and 2\% improvement, respectively. Unlike AOC-IDS and CIDS, which label inputs autonomously, FeCo relies on a fixed threshold in the decision-making, leading to poor adaptability in the online setting. Moreover, the use of cross-entropy loss in the CIDS impairs its representation ability. We further conduct an ablation study by replacing our CRC loss with cross-entropy loss to validate this adverse effect.
        
        It is noteworthy that some traditional machine learning models (DTC, RF, and XGBoost) show poor performance on the NSL-KDD dataset, but relatively good performance on the UNSW-NB15 dataset. This can be attributed to the presence of zero-day attacks in the NSL-KDD dataset, as opposed to the attacks in the UNSW-NB15 dataset which have all previously occurred in the training data. This observation suggests that traditional machine learning models are capable of managing network traffic that is relatively simplistic and does not evolve over time, while more complex deep learning models may be prone to overfitting in these tasks. However, the bad performance of these traditional machine learning-based methods on the NSL-KDD underscores their limitations when applied to more complex and practical scenarios.
        
        
        

\begin{table}[!t]
\caption{Attack distribution of training and test datasets of NSL-KDD. $N_{sample}$ represents the number of samples. $N_{type}$ is the number of attack categories, composed by the sum of two numbers. The first number indicates the attack categories appeared in both the Training \& Test sets, while the second number indicates attack categories appeared only in the Training set / Test set.}
\centering
\begin{tabular}{ccccc}
\toprule
\multirow{2}{*}{Category} & \multicolumn{2}{c}{Training set} &\multicolumn{2}{c}{Test set} \\ 
\cmidrule(lr){2-3} \cmidrule(lr){4-5}

&$N_{sample}$ &$N_{type}$ &$N_{sample}$ &$N_{type}$     \\ 
\midrule
DoS &45927 &6+0 &7458 &6+4 \\
Probe &11656 &4+0 &2421 &4+2  \\
R2L &995 &6+2 &2754 &6+8  \\
U2R &52 &4+0 &200 &4+3  \\
\bottomrule
\end{tabular}
\label{tab:nsl}
\vspace{-0.5cm}
\end{table}

        \begin{figure*}
            \centering
            \includegraphics[width=\textwidth]{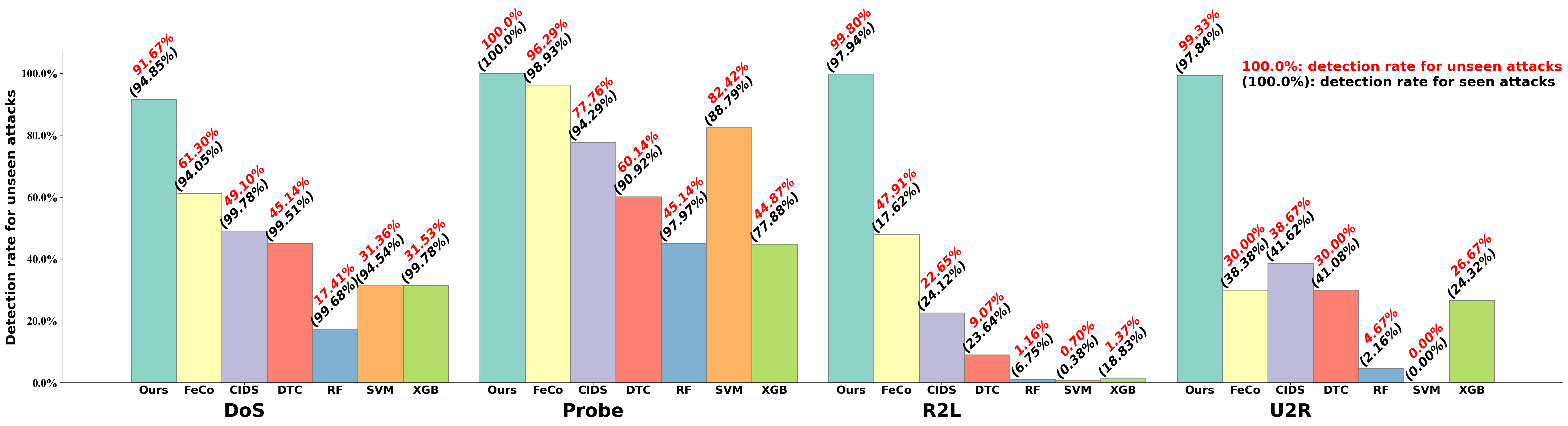}
           \caption{Detection rates of zero-day (unseen) attacks for ours (AOC-IDS) and comparative methods. The detection rate for previously seen attacks is provided in brackets for reference. Numbers in red and black represent detection rates for seen and unseen attacks, respectively.}
           \label{fig:zero_day}
           \vspace{-0.405cm}
        \end{figure*}

    \subsubsection{Detection Rate for Zero-day Attacks}
        To show the detection capacity of IDSs in the dynamic environment, we further analyze the detection for zero-day attacks in the NSL-KDD dataset, whose distribution is detailed in \tablename{~\ref{tab:nsl}}. The test dataset contains zero-day attacks across all four categories of attacks, each representing instances that have not appeared in the training dataset and thus pose new and unique challenges for the detection model. Therefore, each category can be subdivided into `seen' and `unseen' in the training dataset, for instance, `DoS\_seen' and `DoS\_unseen', with all `unseen' attacks considered zero-day for the IDS.
        
        We use the recall for each kind of attack to evaluate the detection rate of each kind of attack, as the recall measures the proportion of attacks that are detected. Given the prediction of $j$-th instance $\hat{y}_j$ and the true label of $j$-th instance $y_j^{label}$, the recall of each attack category $i\in$ \{`attack\_seen',`attack\_unseen'\}, where $attack\in$\{`DoS',`Probe',`U2R',`R2L'\} is defined as 
        
        \begin{equation}
        R(i)=\frac{\sum_j\mathbbm{1}({\hat{y}_j=1 \& y_j^{label}=i})}{\sum_j\mathbbm{1}(y_j^{label}=i)},
        \end{equation}
        where $\hat{y}_j\in\{0,1\}$ is binary results of normal with label `0' and attack with label `1', and $\mathbbm{1}(\cdot)$ is the indicator function, which equals one if the condition in the function is satisfied. 
        
        \figurename{~\ref{fig:zero_day}} demonstrates AOC-IDS surpasses all comparison methods in the detection of zero-day attacks across all categories, with a remarkable detection rate of over 91\%. This high level of performance highlights AOC-IDS's capacity in managing zero-day attack challenges in dynamic environments.

        Despite a few selected methodologies, such as FeCo and SVM, attaining satisfactory detection rates (over 80\%) for zero-day attacks in the Probe category, they fall short in maintaining consistent performance across various attack categories.
        Certain methodologies, including RF and XGBoost, present less than a 50\% recall rate, thereby indicating their inadequacy for detecting zero-day attacks.
        
        Significantly, the performance drop of AOC-IDS between known and unknown attacks is minimal. This contrasts with some methods which, while effective against known attacks, see a severe decline when facing unknown ones. This consistent performance demonstrates the IDS's capacity to accurately define normal behavior patterns by utilizing a limited range of attack types, showcasing its resilience when faced with the unpredictability of zero-day attacks.
        

\subsection{Ablation Experiments}
\label{sec:ablation experiments}

    In this section, we conduct ablation experiments to delineate the efficacy of various components and design strategies in AOC-IDS, including the CRC loss function, the integration of encoder and decoder outputs, and the decision-making process.
    The analysis underscores their collective influence on boosting representational learning capabilities and lessening manual intervention, thereby promoting adaptability and compatibility with dynamic environments.

    \subsubsection{CRC Loss} We compared our custom-designed CRC loss with standard contrastive loss, i.e., InfoNCE loss and cross-entropy loss, denoted by `w/o CRC loss' and `w/o contrastive loss' respectively. In \tablename{~\ref{tab:ablation}}, the result shows the CRC loss outperforms the InfoNCE loss. This verifies the superiority of the dual-category repulsion effect produced by the CRC loss because it avoids the possible repetitive pulling and pushing of negative sample pairs in the standard InfoNCE.
    Moreover, \tablename{~\ref{tab:ablation}} provides evidence that utilizing contrastive losses, which includes our custom-crafted CRC and the standard InfoNCE, outperforms the use of classification (cross-entropy) loss. This underscores the importance of enhanced representation learning capabilities offered by contrastive loss for intrusion detection tasks.
    When cross-entropy loss is employed, the intrusion detection task can be solely handled by the AE, which directly predicts labels for the binary classification task (`normal' or `abnormal'). Conversely, when applying contrastive loss in our experimental setup, the primary role of AE becomes learning representations, rather than predicting directly. Subsequent post-processing is then carried out to determine the predicted label. 
    By breaking down the intrusion detection task into two distinct, more nuanced parts, the detection model is able to focus on the representation learning task resulting in a more powerful representation learning ability brought by contrastive learning. This approach leads to notable performance enhancements compared to systems that rely solely on binary cross-entropy loss for classification.
    
    \subsubsection{Encoder \& Decoder}
    We compared the methods of solely using the output information from either the encoder, described as `w/o decoder', or the decoder, described as `w/o encoder', to train the model in ACO-IDS. 
    \tablename{~\ref{tab:ablation}} suggests that using information from both the encoder and decoder outputs enhances performance. This is due to the comprehensive integration of all available information to learn more insightful representations from the input data, thereby improving the model's learning capability and facilitating a deeper understanding of the input data.

    \subsubsection{Decision Making Process}
    We compared our adaptive, fully automated decision-making process with a fixed threshold method, represented by `w/o DC pro'. In the ablation study with a fixed threshold, after obtaining the cosine similarity score distribution, instead of fitting it into two Gaussian distributions, we set a threshold as $p\%$ lowest normal cosine similarity score as in \cite{FeCo}. Note that the threshold of $p\%$ has specific statistical implications, representing the tolerable false positive rate on the test set. Therefore, it is best to select a statistically common value. In this experiment, $p$ is set to 5. \tablename{~\ref{tab:ablation}} illustrates that adopting a fixed threshold resulted in a decline in system performance. Consequently, our adaptive and fully automated decision-making process is advantageous.

\begin{table}[!t]
\caption{Ablation study results for AOC-IDS. The highest metric performance is bolded.}
\resizebox{\columnwidth}{!}{
\begin{tabular}{ccccccccc}
\toprule
\multirow{2}{*}{Method} & \multicolumn{4}{c}{NSL-KDD} & \multicolumn{4}{c}{UNSW-NB15} \\ \cmidrule(lr){2-5} \cmidrule(lr){6-9}

& Acc.   & Pre.  & Rec.  & F1  & Acc.    & Pre.   & Rec.   & F1   \\ \midrule
AOC-IDS                      & \textbf{88.90}  &85.99 & 96.21 &  \textbf{90.81} & \textbf{89.19}  & \textbf{90.65}   & 89.70    & \textbf{90.14} \\
w/o CRC loss            &   88.79     &   86.87   &    94.62   &  90.56  &   85.88     &   86.66    &  89.44     &   87.43  \\
w/o contrasitve loss    &  74.62 &   \textbf{91.65}   &    60.96   &  73.21   &    81.49    &   75.04    &   \textbf{99.48}    &   85.55   \\
w/o decoder             & 87.68       & 87.28    &  91.90     & 89.44   &  69.19      & 70.56      & 75.38     & 72.79    \\
w/o encoder             & 87.62       & 83.82    &  \textbf{97.06}    & 89.94    & 84.62      & 88.63      & 82.84     & 85.56    \\
w/o DC pro.          &86.02   &89.84      & 85.04      &87.34 & 86.56     & 83.17      & 94.76     & 88.59     \\
\bottomrule
\end{tabular}}
\label{tab:ablation}
\vspace{-0.5cm}
\end{table}

\subsection{Analysis of Online Learning Performance}
    The online setting is specially designed for the dynamic and ever-changing environment in the real world. 
    The two selected datasets represent two distinct scenarios: NSL-KDD approximates real-world conditions with its dynamic network traffic behavior and the potential for zero-day attacks, while UNSW-NB15 presents a more idealized scenario with static network traffic behavior.
    Online learning is much more important in the former than in the latter scenario. 

    Hence, we conducted experiments to emphasize the different importance of online-learning deployment in these two scenarios. The online training process consists of initial and subsequent training phases. Performance post-initial training refers to the system being trained solely based on the initial labeled dataset. As illustrated in \tablename{~\ref{tab:ablation}}, our Intrusion Detection System (IDS) exhibits performance improvements after completing online training in both the NSL-KDD and UNSW-NB15 datasets, with a greater enhancement observed on the NSL-KDD dataset compared to the UNSW-NB15 dataset. 
    This observation validates the compatibility of AOC-IDS in the online setting and implies NSL-KDD, being a more challenging and dynamic dataset, benefits more from the online setting.
    As for UNSW-NB15 with a relatively uniform distribution in the training dataset, the model can achieve satisfactory performance with sufficient training on the limited training dataset, leading to less significant performance improvement after completing online training. 
    In contrast, NSL-KDD is a more intricate dataset, and randomly selected data subsets fail to represent the entire dataset. Consequently, in the online setting, the performance shows greater improvement when exposed to all the input traffic in the training dataset on the NSL-KDD dataset.

    The performance of offline training corresponds to training the system with complete access to the label information of the original training dataset. 
    As illustrated in \tablename{~\ref{tab:ablation}}, in the online setting, there is a slight decrease in performance on the NSL-KDD dataset compared to offline settings. It is attributed to the limited labeled data in the online setting, making it more challenging. 
    Nevertheless, this slight decrease in model performance under online settings is still considered acceptable and does not significantly compromise the efficacy of the IDS in the online setting.
    Regarding the UNSW-NB15 dataset, since the partial training dataset effectively represents the entire training dataset, there is minimal performance variation between online and offline training scenarios.

\begin{table}[!t]
\caption{Performance (\%) of AOC-IDS: comparisons between complete online training, initial online training, and offline training. $\Delta$ denotes the performance difference relative to complete online training. The highest metric performance is bolded.}
\scriptsize
\resizebox{\columnwidth}{!}{%
\begin{tabular}{ccccccccc}
\toprule
\multirow{2}{*}{Setting} & \multicolumn{4}{c}{NSL-KDD} & \multicolumn{4}{c}{UNSW-NB15} \\ \cmidrule(lr){2-5} \cmidrule(lr){6-9} 
  & Acc.   & Pre.   & Rec.  & F1  & Acc.    & Pre.   & Rec.   & F1   \\ \midrule

Online     &   88.90  & 85.99  &\textbf{96.21} &  90.81 & 89.19  & 90.65   & \textbf{89.70}    & \textbf{90.14}      \\
\cmidrule{1-9}
Initial      &   85.92  &   88.14  &  87.09 &  87.46 & 88.70     & 96.64    & 82.45     & 88.91       \\
  $\Delta$   &   +2.98 & -2.15 & +9.12 & +3.35 & +0.46 & -5.99 & +7.25 & +1.23\\
\cmidrule{1-9}
Offline     &   \textbf{91.02}       & \textbf{90.11}      & 95.38       &   \textbf{92.36}    &\textbf{89.67}        &\textbf{99.44}       &85.81       & 90.12         \\
  $\Delta$   &  -2.12 & -4.12 & +0.83 & -1.55 & -0.48 & -8.79 & +3.89 & +0.02\\
\bottomrule
\end{tabular}}
\label{tab:compare_online}
\vspace{-0.5cm}
\end{table}

\section{Related Work}
In this section, we first discuss the existing works on intrusion detection. Then, we introduce online learning and contrastive learning including their applications in intrusion detection, to explain the research gap.

\subsection{Intrusion Detection}
Intrusion detection refers to the process of identifying and responding to unauthorized or malicious activities that pose a threat to system security \cite{habeeb2022network}. 
The development of machine learning, especially deep learning, has significantly advanced intrusion detection research, since they are effective and efficient in learning complex representations \cite{SARSA, FERRAG2020102419}. 

Various traditional machine learning methods, such as SVM and random forest \cite{5687239, 8369054}, and classic deep learning architectures, such as convolutional neural network (CNN) \cite{8126009,8066291} and recurrent neural network (RNN) have been applied to intrusion detection. Ferrag et al. \cite{FERRAG2020102419} reviewed the works of intrusion detection systems based on deep learning approaches and analyzed the performance of different deep learning models, including RNNs, CNNs, deep AEs, restricted Boltzmann machines, deep Boltzmann machines, and deep belief networks. 

However, intrusion detection still faces several challenges, including adaptability to new system patterns, zero-day attacks, and labor-intensive labeling. To address these issues, online learning and contrastive learning have been incorporated into IDS.

\subsection{Online Learning}

Online learning, proposed to address challenges in dynamic environments, is a machine learning paradigm that involves periodically updating a model as new data becomes available \cite{wahab2022intrusion}. This approach enables the model to adapt to changing data distributions and facilitates continuous updates \cite{yang2021lightweight,wahab2022intrusion}. As a result, online learning is well-suited for various applications, such as recommendation systems \cite{xiao2018personalized}, fraud detection \cite{dornadula2019credit}, spam filtering \cite{wang2006svm}, resource allocation \cite{jiang2021crowdpatrol}, and other domains where adaptive decision-making is crucial.


In the context of IoT, the utilization of the online learning paradigm has led to significant achievements due to its adaptability \cite{chen2018heterogeneous}. Although many online frameworks for IDS can update the models periodically to adapt to dynamic environments, they require labor-intensive labeling for continuous training \cite{han2021log}. To address this issue, Han et al. \cite{hananomaly} proposed an online deep-learning framework that only labels the most influential samples to reduce the labeling overhead. Gyamfi et al. \cite{gyamfi2022novel} developed a novel online network intrusion detection system that can update the model when the environment uncovers the dynamic attacks. Xu et al. \cite{xu2020method} proposed a few-shot meta-learning framework for IDS with only a limited number of shots of malicious samples. 

In this work, we propose an autonomous online-learning framework compatible with inherent streaming data in IoT systems. This framework facilitates dynamic system updating without the need for laborious manual labeling, thereby augmenting its capability to manage evolving data streams.






\subsection{Contrastive Learning}
Contrastive learning learns representations by contrasting positive samples with negative samples \cite{tian2020makes}. The goal is to enhance the similarity between similar samples and increase the dissimilarity between dissimilar samples in the learned representation space \cite{khosla2020supervised}. 
InfoNCE \cite{InfoNCE} is the most commonly used loss function in contrastive learning, measuring the similarity between two views by comparing their representations \cite{hoffmann2022ranking}. 
Contrastive learning has achieved state-of-the-art results in various domains, including computer vision \cite{dai2017contrastive}, natural language processing \cite{le2020contrastive}, and mobile computing \cite{zhou2023mobile}. 

Contrastive learning has also been applied to enhance the performance of intrusion detection systems. For instance, Wang et al. \cite{FeCo} defined a contrastive loss for maximizing the distance between benign and malicious samples and at the same time, minimizing the distance among benign samples. In \cite{conFourier}, a contrastive learning scheme was proposed to detect both known and unknown attacks, expanding the representation capacity of intrusion detection systems and outperforming similar models in detecting previously unseen attacks. Yue et al. \cite{conIDS} introduced a contrastive learning-based method for network intrusion detection, which improved accuracy while reducing false detections caused by intra-class diversity and inter-class similarity. Liu et al. \cite{9395172} proposed CoLA, a contrastive learning model that leveraged local information from network data for anomaly detection in attributed networks. 

In this study, we design a novel contrastive loss, specifically tailored to intrusion detection tasks. Operating within a binary classification paradigm, this function distinguishes between two categories – `normal' and `abnormal', and generates a dual-category repulsion effect at each step.

\section{Conclusion}


In this paper, we introduce AOC-IDS, designed for intrusion detection in dynamic network environments with evolving attacks and system behaviors. 
By integrating an AE equipped with a tailored CRC loss function, which utilizes the outputs of both the encoder and decoder, AOC-IDS achieves superior data interpretation and enhanced representational capacity. These elements facilitate the more effective differentiation of malicious activities from standard system patterns. 
Crucially, the online learning framework facilitates autonomous continual adaptation of the system with the help of pseudo-labels automated generated from the ADM. The elimination of manual labeling bolsters the practicality of AOC-IDS. 
We substantiate the exceptional performance and adaptability of AOC-IDS through comparative experiments on two datasets: NSL-KDD and UNSW-NB15. Our ablation study provides an insightful analysis of the contributions of each system component. Further analysis confirms that model updating indeed enhances overall performance. When compared to offline learning, our system maintains performance with a significant reduction in the volume of information required. 

\section*{Acknowledgment}
This research is supported in part by the National Natural Science Foundation of China (Grant No. 92067109, 61873119, 62211530106), in part by the Shenzhen Science and Technology Program (Grant No. ZDSYS20210623092007023, GJHZ20210705141808024), in part by the Educational Commission of Guangdong Province (Grant No. 2019KZDZX1018), and in part by the UGC General Research Fund (Grant No. 17203320, 17209822) from Hong Kong.

\bibliographystyle{IEEEtran}
\bibliography{references}

\end{document}